\RequirePackage{fix-cm}
\documentclass[smallextended]{svjour3}       % onecolumn (second format)
\smartqed  % flush right qed marks, e.g. at end of proof
\usepackage{graphicx}
\usepackage{setspace}

\def\bSig\mathbf{\Sigma}

 \newcommand{\bbeta}{\mbox{\boldmath $\beta$}}
 \newcommand{\bdZ}{\mbox{\boldmath $Z$}}

\begin{document}

\title{Matching methods for obtaining survival functions to estimate the effect of a time-dependent treatment\thanks{This work was supported in part by National Institutes of Health grant 5R01-DK070869.
The authors thank the Canadian Institute for Health Information for
access to the Canadian Organ Replacement Register database.}
}
%\subtitle{Matching methods to estimate time-dependent treatment effect}

\titlerunning{Matching methods to estimate time-dependent treatment effect}        % if too long for running head

\author{Yun Li         \and
        Douglas E. Schaubel \and Kevin He
}


\institute{Y. Li \at
             Department of Biostatistics, University of Michigan, Ann Arbor, MI, 48109-2029, USA \\ Division of Biostatistics, University of Pennsylvania, Philadelphia, PA, 19104-6021. 
              \email{yun.li@pennmedicine.upenn.edu}           %  \\
           \and
           D. Schaubel and K. He   \at
              Department of Biostatistics, University of Michigan, Ann Arbor, MI, 48109-2029, USA
}

\date{Received: date / Accepted: date}

\maketitle

\begin{abstract}
In observational studies of survival time featuring a binary time-dependent treatment, the hazard ratio (an instantaneous measure) is often used to represent the treatment effect. However, investigators are often more interested in the difference in survival functions. We propose semiparametric methods to estimate the causal effect of treatment among the treated with respect to survival probability. The objective is to compare post-treatment survival with the survival function that would have been observed in the absence of treatment.
For each patient, we compute a prognostic score (based on the pre-treatment death hazard) and a propensity score
(based on the treatment hazard). Each treated patient is then matched with an alive, uncensored and not-yet-treated patient with
similar prognostic and/or propensity scores. The experience of each treated and matched patient is weighted using a variant of Inverse Probability of Censoring Weighting to account for the impact of censoring. We propose estimators of the treatment-specific survival functions (and their difference), computed through weighted Nelson-Aalen estimators. Closed-form variance estimators are proposed which take into consideration the potential replication of subjects across matched sets. The proposed methods are evaluated through simulation, then applied to estimate the effect of kidney transplantation on survival among end-stage renal disease patients using data from a national organ failure registry.
\end{abstract}

\section{Introduction}
\label{s:intro}
For medical studies in which time to a failure event is of interest, the effect of a treatment is often estimated by comparing the survival functions for the treated and untreated groups. When treatment is assigned at baseline (time $t=0$), the estimation of the survival functions is usually straightforward. In our setting, treatment assignment is time-dependent and a stochastic process such that subjects typically begin follow-up untreated, with some going on to receive treatment at some time after baseline. In this report, we are primarily considering observational studies in which treatment is not assigned at random and the rate of treatment assignment may depend strongly on follow-up time and covariates ($\bdZ$). Of chief interest is to estimate the average effect of treatment on the treated (ATT), for the purposes of providing a summary evaluation of the impact of the treatment under the existing assignment patterns. The ATT is a useful alternative to the average causal effect (ACE), and may be more relevant and preferred in various settings (including ours) with respect to policy implications (Heckman et al. 1997; Schafer et al. 2008).

Methods proposed in this report are motivated by the objective of estimating the effect of deceased-donor kidney transplantation (treatment) compared to dialysis (``untreated'') on the survival function. End-stage renal disease patients typically begin therapy on dialysis, with some later receiving a kidney transplant. The referral of patients for kidney transplantation is not random, as only patients deemed medically
suitable are considered.  The goal is to estimate the average effect on the survival function of kidney transplantation under
current transplant referral practices; i.e., under the current set of decisions influencing which (and when during follow-up) patients tend to get transplanted.  The ATT, in this context, is intended to contrast the average post-transplant survival function with the average survival function that
would have been observed (among the transplanted patients) had kidney transplantation not been available.

The effect of a time-dependent treatment is often evaluated using Cox regression, with treatment receipt (yes/no) represented by a time-dependent indicator. From such a model, the treatment effect is usually summarized by the hazard ratio. However, investigators are often more interested in contrasting survival (as opposed to hazard) functions, for several reasons. First, the survival function is more interpretable to non-statisticians than the hazard function.  Second, contrasts between survival functions reflect the cumulative effect of treatment, rather than instantaneous treatment effect estimated through the hazard ratio. The hazard ratio estimates the cumulative treatment effect only under proportionality between the pre- and post-treatment hazard functions, which one would often prefer not to assume.  For example, in the motivating example described above, non-proportionality of the pre- and post-kidney-transplant mortality hazards has been reported in the nephology literature for more than 10 years (e.g., Wolfe et al. 1999).

Many methods are available in the existing literature for evaluating the effect of a time-dependent treatment from observational data, as summarized by Robins and Hern$\mbox{\'{a}}$n (2009).  However, most existing methods do not target the ATT specifically in terms of the survival functions. Marginal structural models (Robins et al. 2000; Hern$\mbox{\'{a}}$n et al. 2000; Hern$\mbox{\'{a}}$n et al. 2001) and their history-adjusted versions (e.g., Petersen et al. 2007) typically estimate the causal hazard ratio (HR) as a measure of the ACE of treatment.  When fitted though g-estimation (Robins et al. 1992; Lok et al. 2004; Hern$\mbox{\'{a}}$n et al. 2005), structural nested failure time models (SNFTMs) often use the accelerated failure time model as the basis for the time-dependent treatment effect, in which case mean survival times are contrasted, but not survival functions. When fitted through parametric g-computation (Robins, 1986, 1987 and 1988; Taubman et al. 2009), SNFTMs could in principle be used to estimate the survival function-based ATT of interest in the current report. Disadvantages of such an approach include the need to bootstrap and greater sensitivity towards any model misspecification (Taubman et al. 2009).  Further comparison between the proposed methods and existing approaches is deferred to Section 5.

When treatment is time-dependent, it is generally not straightforward to compare the average post-treatment and treatment-free survival functions.
Several authors have advocated landmark methods (e.g., Feuer et al. 1992). However, the selection of the landmark times at which to classify patient treatment status is arbitrary.  The methods of Feuer et al (1992) were not designed to incorporate covariates. More recent related work includes that of Van Houwelingen (2007) and Van Houwelingen and Putter (2007), which accommodated covariates but did not consider the average treatment effect.

In this paper, we propose matching methods to estimate ATT. Compared with alternative methods, the matching methods have the advantages of handling covariates of higher dimensions, greater robustness towards model misspecifications and less stringency towards positivity assumptions (Rosenbaum and Rubin, 1983). Furthermore, matching can be more intuitive to researchers and does not rely on structural models. We propose to select matches to serve as potential treatment-free counterfactuals (or controls) for treated patients. Specifically, for a patient initiating treatment at a certain follow-up time (e.g., time $T$), we propose to select his match from patients alive, uncensored and not-yet-treated at time $T$. The matched patient is intended to be very similar to the treated patient, such that their follow-up (after time $T$) reflects what would have been the treated patient's experience, had (contrary to fact) that patient not been treated. We consider 1:1 matching, which equalizes the follow-up time distribution (i.e., previous time survived) prior to time $T$ between the treated and matched yet-untreated subject. We consider two scores by which to match patients: (1) a propensity score which measures the patient-specific rate of treatment assignment, given the covariates (2) a prognostic score which represents the pre-treatment death hazard. Hence, such matching balances the covariate distribution by requiring the matched patient to be very similar to the treated patient with respect to the rate of receiving treatment and/or the rate of dying in the absence of treatment. After appropriate reweighting, group-specific survival curves are then estimated and compared nonparametrically such that no functional form for the treatment effect is assumed. We target at ATT and the time-dependent treatment of interest is non-reversible. It is an important special case of time-varying treatment regimes and the estimation process through matching raises non-trivial technical challenges.

Several complications arise from censoring that have the potential to bias a survival function estimator. First, the treatment time ($T$) is subject to censoring, such that longer times-to-treatment are more likely to be censored. Hence, the observed distribution of $T$ is generally a biased sample of shorter $T$ values. Second, the treatment time $T$ and the post-treatment death time ($D-T$) are not usually independent. Thus, $(D-T)$ is inherently subject to dependent censoring, a phenomenon referred to in the gap time literature as induced dependent censoring (e.g., Lin, Sun and Ying 1999; Schaubel and Cai 2004).  Third, matched yet-untreated patients can later receive treatment after being matched. To eliminate these sources of biases, we weight the estimators using a variant of Inverse Probability of Censoring Weighting (IPCW; Robins and Rotnitzky 1992; Robins and Finkelstein 2000) in order to recover the survival and time-to-treatment distributions that would be observed in the absence of censoring.

The remainder of this article is organized as follows. In Section 2, we will describe our proposed methods, including the pertinent counterfactuals, matching design and assumed models, proposed treatment effect and variance estimators. In Section 3, we conduct simulations to evaluate the performance of the survival estimators and the associated variance estimators.  Furthermore, we evaluate the bias and efficiency of the survival estimators when one or both scores are used to conduct matching. In Section 4, we apply the methods to national end-stage renal disease data. We conclude the paper with some discussion in Section 5.

\section{Methods}
\label{s:model}
In this section, we define the quantity of interest, then describe the criteria and process to select matches for each treated patient.
 We then address the issue of censoring and the need to weight the analysis. Next, we introduce our proposed estimators of the post-treatment and treatment-free survival functions, and the difference therein; after which, variance estimation is outlined.

\subsection{Notation, Quantity of Interest and Identifiability}
We begin by setting up the requisite notation. Let $D_i$ denote the death time and $C_i$ the censoring time for subject $i$ ($i=1,\ldots,n$). The observation time is denoted by $U_i = D_i \wedge C_i$, with $a\wedge b=\min\{a,b\}$, and the death indicator is given by $\Delta_i = I(D_i < C_i)$ where $I(A)$ is an indicator function taking the value 1 when event $A$ is true and 0 otherwise.  The at-risk indicator is defined as $Y_i(t) = I(U_i \geq t)$.  Let $\bdZ_i$ be the covariate vector, which is assumed to not depend on time. The treatment time is represented by $T_i$, with corresponding indicator $\Delta_i^T=I(T_i < U_i )$.
We assume that, conditional on $\bdZ_i$, $D_i$ and $T_i$ are independently censored by $C_i$. A few comments are in order regarding our data structure. As implied previously, patients begin follow-up ($t=0$) untreated, with some subsequently receiving treatment and others dying first. Treatment does not censor death, but does naturally preclude treatment-free death. Correspondingly, death prevents future treatment initiation, as in the competing risks setting.

As stated in the preceding paragraph, we assume that the covariate, $\bdZ_i$, does not vary with time. Several of the methods cited in Section 1 are able to accommodate time-varying covariates.  Three ideas are important in this regard. First, with respect to the proposed methods, the innovation relates to the methods of estimation and (to some extent) the estimands themselves, as opposed to the underlying data structure. Second, data sets with time-constant covariates are common in practice; such as administrative databases, including that which motivates our current work.  Third, it appears that the proposed methods could be extended to the time-dependent covariate setting with little modification. These are issues we return to in Section 5.

We define the parameter of interest in the causal inference framework. Typically, this framework hypothesizes the setting wherein each individual has two potential outcomes (Rubin 1974 and 1978), corresponding to the two possible treatment regimes (e.g., treated and untreated). We modify this structure to accommodate
our setting. Let $D_i^1(T_i)$ denote the potential death time (measured from time 0) if patient $i$ is treated at $T_i$. The counterfactual quantity $D_i^0(T_i)$ denotes the potential death time if, contrary to fact, patient $i$ never received treatment. Note that, by definition, both $D_i^0(T_i)$ and $D_i^1(T_i)$ are greater than $T_i$ and the counterfactuals are only defined in individuals that begin treatment. In the absence of censoring, $D_i^1(T_i)=D_i$. We assume that $D_i^1(T_i)$ and $D_i^0(T_i)$ are conditionally independent of the treatment assignment given the observed covariates, known as the strong ignorability assumption (Rubin, 1974). We also assume the stable unit treatment value assumption (Rubin 1980).

Our objective is to estimate the average effect of the treatment among the treated, a frequently employed measure in the causal inference literature. The treatment decision depends on $\bdZ_i$ and untreated patients may never be eligible for treatment. Additionally, the benefit of treatment is only realized among patients who actually receive treatment. Hence, the ATT can be more desirable in practice than the average causal effect (ACE). For patient $i$, let $\widetilde{D}_i^1(T_i)$ denote the potential remaining survival time following treatment assignment at $T_i$, such that $\widetilde{D}_i^1(T_i)=D^1_i(T_i) - T_i$.  Conversely, let $\widetilde{D}_i^0(T_i)$ denote the potential remaining survival time if the patient never receives treatment; i.e., $\widetilde{D}_i^0(T_i)=D^0_i(T_i) - T_i$. In the absence of censoring, the survival functions
corresponding to these newly defined variates are given by,
\begin{eqnarray}
S_{ij}(t) & = & P\{\widetilde{D}_i^j(T_i) > t |T_i,\bdZ_i,T_i<D_i\},  \;\;\;\;\;\;  j=0,1 \nonumber
\end{eqnarray}
and the subject-specific treatment effect can be defined as
\begin{eqnarray}
\delta_i(t) =  S_{i1}(t) - S_{i0}(t).
\nonumber
\end{eqnarray}
Having described the treatment effect at the individual level, we now denote the
average causal treatment effect among the treated as
\begin{eqnarray}
\delta(t) =  S_1(t) - S_0(t), \label{eq:delta}
\end{eqnarray}
where $S_0(t)$ and $S_1(t)$ are average survival functions,
\begin{eqnarray}
S_j(t)=  E\{S_{ij}(t)\} \label{eq:sj}
\end{eqnarray}
with the expectation being with respect to the distribution of $\{T,\bdZ|T<D\}$; i.e., the joint distribution of $(T,Z)$ among patients with $T<D$,
which accounts for the competing risks relationship between $T$ and $D$. In addition to our inherent interest in the ATT, it should be noted that
what makes the estimation of the ATT more feasible than the ACE in our setting is that we only observe the pre-treatment duration ($T$) for subjects observed to receive treatment.

It is important to understand which quantities pertinent to estimating the ATT can be identified by observed data. In the absence of censoring, we would observe $\widetilde{D}^1(T)$ for each treated patient and, hence, could estimate $P\{\widetilde{D}^1(T) > t |\bdZ, T, T<D\}$ which is equal to $P\{(D - T) > t |\bdZ, T, T<D\}$ under the strong ignorability assumption. However, we do not observe data to estimate $P\{\widetilde{D}^0(T) > t |\bdZ, T, T<D\}$ since a subject's treatment-free experience is censored at the time of treatment.  Therefore, we use matching methods to choose proper substitutions from the alive, uncensored and not-yet-treated patients. Specifically, we achieve this by hard-matching on $T$ and matching on $\bdZ$ using a prognostic score and/or propensity score, $p(\bdZ)$, to select closest matches to serve as treatment-free counterfactuals for each treated patient. To estimate the ATT, instead of averaging over the conditional distribution, $(T,\bdZ|T<D)$, we average over the distribution of $\{T,p(\bdZ)|T<D\}$ and apply the result that the potential remaining survival times after $T$ are conditionally independent of the treatment assignment given the matching scores (Rosenbaum and Rubin 1983; Lu 2005). We also assume that subjects with the same $\bdZ$ have a positive probability of being in both treatment groups (Heckman, LaLonde and Smith 1999) or an overlapping support for treated and untreated patients. Additionally, we assume no unmeasured confounders or measurement errors or model misspecifications. See related discussions on assumptions required for matching (Bryson, Dorsett and Purdone 2002; Caliendo and Kopeinig 2005; Rosenbaum and Rubin 1983; Stuart 2010). In the presence of censoring, instead of estimating $P\{\widetilde{D}^j(T) > t |\bdZ, T,T<D\}$, we can estimate $P\{\widetilde{D}^j(T) > t |\bdZ, T, \Delta^T=1\}$, assuming that censoring is conditionally independent given $\bdZ$. The implications of censoring will be addressed in details in Subsections 2.3 and 2.4.

In the next subsection, we describe how to select matches for treated patients in order to estimate $P\{\widetilde{D}^0(T) > t |\bdZ, T, \Delta^T=1\}$.

\subsection{Matching schemes}
We consider two scores as potential matching criteria. The first is a propensity score, based on the cause-specific hazard of initiating treatment at time $t$ given alive and untreated,
\begin{eqnarray}
    \lambda_{iT}(t) = \lim_{dt\rightarrow 0} \frac{1}{dt}P(t \leq T_i\wedge D_i < t+dt, T_i<D_i | T_i\wedge D_i \geq t, \bdZ_i ), \nonumber
\end{eqnarray}
for which we assume the following Cox model,
\begin{eqnarray}
    \lambda_{iT}(t) = \lambda_{0T}(t)\exp(\bbeta_T' \bdZ_i),
\end{eqnarray}
where $\lambda_{0T}(t)$ is an unspecified baseline hazard and $\bbeta_T$ is a vector of unknown parameters. Cox regression is used due to its familiarity and flexibility. For each treated patient, we select matches among at-risk and not-yet-treated patients. For example, consider finding matches for patient $k$, who is treated at time $T_k$. Potential controls for patient $k$ are patients who are alive, uncensored and untreated as of time $t=T_k$, including subjects who are later treated at time $t>T_k$. We compare treated patient, $k$, and a potential control, $\ell$, with respect to treatment propensity through the ratio
\begin{eqnarray}
\psi_{\ell:k}^T \equiv    \frac{\lambda_{l T}(T_k)}{\lambda_{k T}(T_k)} & = & \frac{\lambda_{0T}(T_k)\exp(\bbeta_T'\bdZ_{\ell})}{\lambda_{0T}(T_k)\exp(\bbeta_T'\bdZ_{k})} = \exp\{\bbeta_T'(\bdZ_{\ell} - \bdZ_k)\}.
\end{eqnarray}
The fact that the baseline hazard cancels out simplifies computation considerably. Patient $\ell$ is a suitable match to treated patient $k$ to
the extent that $\psi_{\ell:k}^T$ is close to 1. To avoid inappropriate matches, we add the restriction that the $\psi_{\ell:k}^T$ needs to be within a caliper, $\psi_{\ell:k}^T \in (\xi_T^{-1}, \xi_T)$ for small $\xi_T$. The finite sample positivity assumption and the overlapping common support will impact the selection of $\xi_T$. We select as a match for treated patient $k$ the patient $\ell$ with the minimum $|\log \psi_{\ell:k}^T|$ among at-risk and untreated patients.
It is possible that a treated patient cannot be matched to any control, and such patients are excluded from further analysis.

A second score to be used to choose matches instead of (or, in addition to) $\psi_{\ell:k}^T$ is a prognostic score, reflecting the treatment-free death hazard,
\begin{eqnarray}
    \lambda_{iD}(t) = \lim_{\delta\rightarrow 0} \frac{1}{\delta}P(t \leq D_i < t+\delta |D_i \geq t,T_i > t ,\bdZ_i ), \nonumber
\end{eqnarray}
which we model by
\begin{eqnarray}
    \lambda_{iD}(t) = \lambda_{0D}(t)\exp(\bbeta_D' \bdZ_i).
\end{eqnarray}
To fit this model using standard partial likelihood (Cox 1975) methods, we assume that both $T_i$ and $C_i$ independently censor $D_i^0(T_i)$ given $\bdZ_i$.
Analogous to our use of $\psi_{\ell:k}^T$ described above, we define
\begin{eqnarray}
\psi_{\ell:k}^D  & = & \exp\{\bbeta_D'(\bdZ_{\ell} - Z_k)\},
\end{eqnarray}
then select the match to treated patient $k$ the at-risk and yet-untreated
patient $\ell$ for which $|\log \psi_{\ell:k}^D|$ is minimized, provided that $\psi_{\ell:k}^D \in (\xi_D^{-1}, \xi_D)$ for small $\xi_D$.
We also consider the simultaneous use of both propensity and prognostic scores, in which case the matching requirement would be that $\psi_{\ell:k}^T\in [\xi_T^{-1}, \xi_T]$, $\psi_{\ell:k}^D\in [\xi_D^{-1}, \xi_D]$,
and that $|\log \{\psi_{\ell:k}^T\psi_{\ell:k}^D\}|$ is the minimum among at-risk patients yet-untreated as of time $t=T_k$.

The matching algorithm we propose entails matching-with-replacement in the sense that a patient, while untreated, can be matched to multiple treated patients. Hence, each treated patient is matched to its nearest neighbor within a caliper, even if that subject has been matched to other treated patients at previous times. Figure $1$ serves as an illustration of the matching scheme with $n=4$ hypothetical patients. Following the cohort from time $t=0$ forward, patient $i=2$ receives treatment at time $t=T_2$, and patients $i=1, 3, 4$ are all potential matches. The next observed treatment time is $t=T_1$, with patient $i=3$ being the only potential match. Hence, patient $i=1$ is both a potential control for patient $i=2$ at $t=T_2$ and subsequently a treated patient at $t=T_1$. Patient $i=3$ is a potential match for $i=2$ at $t=T_2$ and then again for $i=1$ at $t=T_1$.
Note that, if patient $i=1$ was selected as a match at $t=T_2$, then $D_1$ does not count as a treatment-free death in this matched set, since $i=1$ is censored upon treatment initiation at time $t=T_1$ in this matched set.

\begin{figure}
\centering
\includegraphics[width=12.9cm, height=20cm]{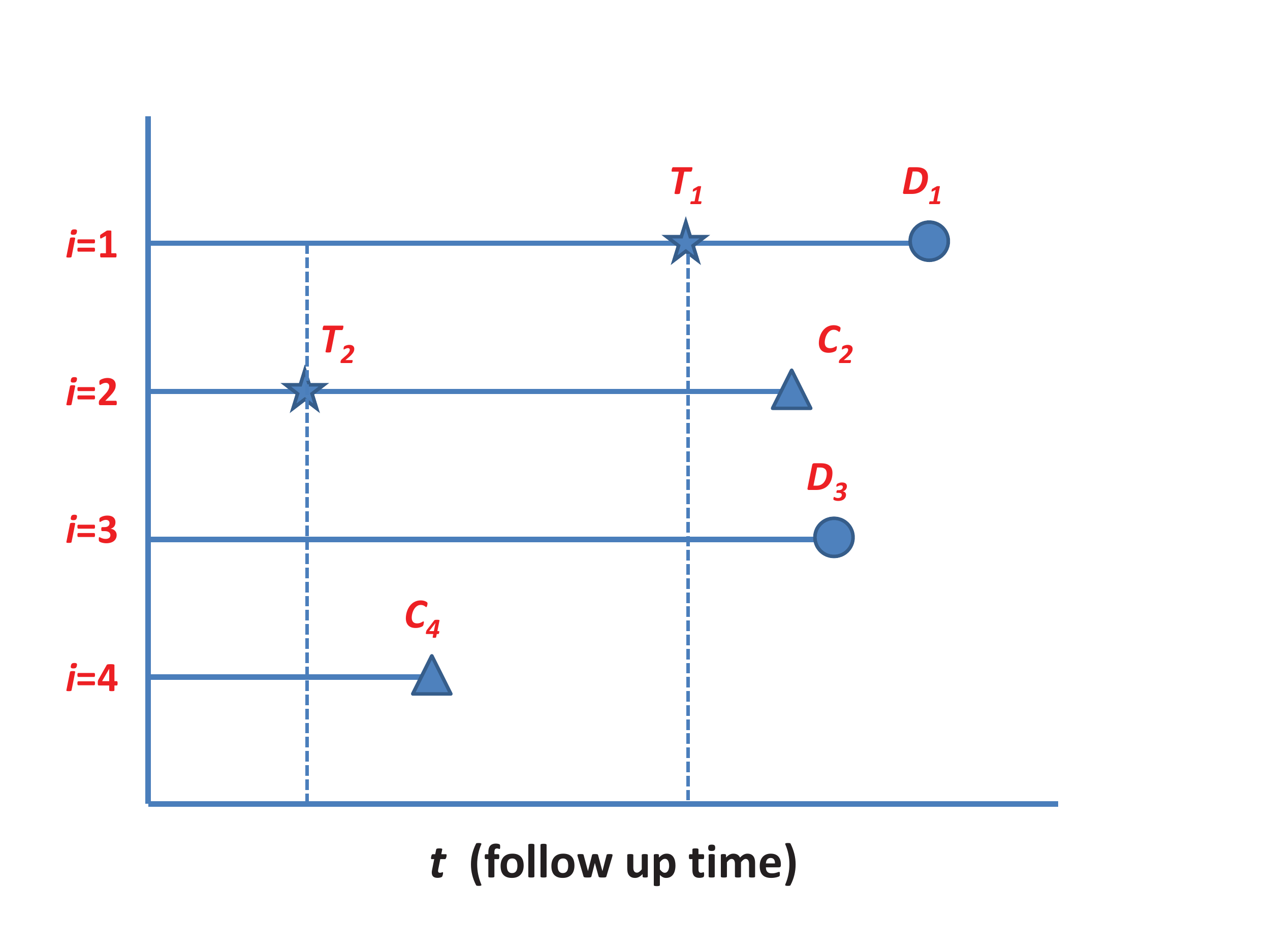}
  \caption{Matching Scheme Illustration}
   \label{fig:1}
\end{figure}

Having described how to select matches for the treated patients, we next describe how to estimate $S_1(t)$ and $S_0(t)$ defined by (\ref{eq:sj}), as well
as the quantity of chief interest, $\delta(t)$ defined by (\ref{eq:delta}).

\subsection{Estimation of $S_1(t)$}

Having created treated and untreated samples that are matched with respect to previous time survived and $\bdZ$, it is appealing to estimate
$S_1(t)$ and $S_0(t)$ nonparametrically.  Recall that the at-risk indicator is defined as $Y_i(t)=I(U_i\geq t)$. The observed death counting process is set to $N_i(t)=\Delta_iI(U_i\leq t)$, with associated increment $dN_i(t)=N_i(t^-+dt)-N_i(t^-)$. In addition, we define the treatment counting process as $N_i^T(t)=\Delta_i^TI(T_i\leq t)$.
We also define $Y_i^1(t)=I(U_i \geq t, T_i<t)$, which equals 1 when subject $i$ is at risk at time $t$ and has already initiated treatment.
Correspondingly, we define the post-treatment counting process increment, $dN_i^1(t)=Y_i^1(t)dN_i(t)$.  Recall that we are analyzing time since treatment initiation, essentially re-setting the time clock for treated patient, $k$, to 0 at the time of treatment, $T_k$. It is then convenient to establish notation that captures the transformed time scale, including
$\widetilde{Y}_k^1(t)=Y_k(T_k+t)$ and $d\widetilde{N}_k^1(t)=dN_k^1(T_k+t)$.

One might hope that this could be accomplished through, for example, the Nelson-Aalen estimator of the cumulative hazard function,
\begin{eqnarray}
\sum_{k=1}^n \int_0^t \left\{ \sum_{\ell=1}^n \widetilde{Y}_\ell^1(u) \right\}^{-1} d\widetilde{N}_k^1(u).
\label{eq:NA}
\end{eqnarray}
However, as we describe below, this turns out not to be the case for several reasons.  First, (\ref{eq:NA}) represents an unweighted average over the observed treatment times.  Since $T_k$ is subject to right censoring by $C_k$, the uncensored $T_k$ values represent a biased sample of shorter values of times-to-treatment.  Just as one would not use the empirical cumulative distribution function for estimation in the standard univariate survival set-up, a method that explicitly accounts for censoring is required here so that the resulting nonparametric estimator of $S_1(t)$ represents an average over the $\{T,\bdZ|T<D\}$ distribution, as opposed to a distribution that depends on $C$.

Second, the length of $T$ affects the probability that $(D-T)$ is observed as opposed to being censored. That is, since we assume one single censoring time, $(D-T)$ is censored by $(C-T)$ which induces dependent censoring (Schaubel and Cai 2004) unless $(D-T)$ is independent of $T$, usually an unrealistic assumption in practice.  Viewing $(D-T)$ as a gap time brings forth an identifiability issue (Lin, Sun and Ying 1999). Specifically, if we let $\tau_C$ be the maximum censoring time, then inference is restricted to $T\in [0,\tau]$ with $S_1(t)$ estimable on $t\in [0,\tau_1]$ for $\tau+\tau_1 \leq \tau_C$.

Third, even leaving aside the first and second issues described above, $C$ still causes difficulty. For instance, suppose that no $T$ values
were censored; that $(D-T)$ is independent of $T$; but that post-treatment death time, $(D-T)$ is still subject to right censoring. The variates $(D-T)$ and $(C-T)$ are likely to be correlated through their mutual association with $\bdZ$, akin to the type of dependent censoring described by Robins and Rotnitzky (1992).

Both the first and third issues described in the preceding paragraphs imply re-weighting the post-treatment data
to reflect that which would be observed in the absence of censoring.  The second issue implies redefining the post-treatment survival function as $S_1(t|T\leq \tau)$ for $t\in [0,\tau_1]$.  Combining these considerations, the post-treatment weight function for subject $k$ is then given by
\begin{eqnarray}
w_k^1(t) & = & \frac{N_k^T(\tau)I_{\bullet:k}\widetilde{Y}_k^1(t)}{P(C_k>T_k+t|\bdZ_k,T_k)},
\label{eq:w1k}
\end{eqnarray}
where $N_k^T(\tau)$ reflects the above-described identifiability constraints, $I_{\bullet:k}$ is an indicator
for treated patient $k$ being successfully matched and $P(C_k>T_k+t|\bdZ_k,T_k)$ is the IPCW component applied so that the
observed data reflect what would have been observed in the absence of censoring.  It is useful to write,
\begin{eqnarray}
P(C_k>T_k+t|\bdZ_k,T_k) & = & P(C_k>T_k|\bdZ_k,T_k) \; P(C_k>T_k+t|C_k>T_k,\bdZ_k,T_k),
\nonumber
\end{eqnarray}
where the first term on the right side represents the probability that treatment time for subject $k$ is uncensored;
while the second term represents the probability the post-treatment death time is uncensored as of
$t$ units following the uncensored treatment time.  All instances of the risk set indicator in the unweighted estimator (\ref{eq:NA}) including that in $d\widetilde{N}_k^1(u)$ will be replaced with $w_k^1(t)$, noting that the counting process increment in (\ref{eq:NA}) is by definition equal to $\widetilde{Y}_k(t)d\widetilde{N}_k^1(u)$.

It is evident now that we require a model for $C_i$, and we assume that
\begin{eqnarray}
    \lambda_{iC}(t) = \lim_{dt\rightarrow 0} \frac{1}{dt}P(t \leq C_i < t+dt |C_i \geq t,\bdZ_i ), \nonumber
\end{eqnarray}
follows the Cox model,
\begin{eqnarray}
    \lambda_{iC}(t) = \lambda_{0C}(t)\exp(\bbeta_C' \bdZ_i).
\end{eqnarray}
Recall that $C_i$ is assumed to be conditionally independent of $T_i$ and $D_i$ given $\bdZ_i$, meaning that
$\Lambda_{0C}(t)=\int_0^t \lambda_{0C}(u)du $ and $\bbeta_C$ can be consistently estimated through standard (unweighted) Cox regression.

Finally, our proposed estimator of $S_1(t)$ is given by $\widehat{S}_1(t)=\exp\{-\widehat{\Lambda}_1(t)   \}$, where
\begin{eqnarray}
\widehat{\Lambda}_1(t) = \sum_{k=1}^n \int_0^t \left\{ \sum_{
\ell=1}^n \widehat{w}_{\ell}^1(u) \right\}^{-1} \widehat{w}_\ell^1(u)d\widetilde{N}_k^1(u)
\end{eqnarray}
and we set
\begin{eqnarray}
\widehat{w}_k^1(t) & = &
N_k^T(\tau)I_{\bullet:k}\widetilde{Y}_k^1(t)
\exp\left\{ \widehat{\Lambda}_{kC}(T_k+t) \right\},
\label{eq:wk1hat}
\end{eqnarray}
where $\widehat{\Lambda}_{kC}(T_k+t)=\int_0^{T_k+t} \widehat{\lambda}_{kC}(u)du $.

Considerations on the treatment-free side are somewhat different from those outlined in this subsection, as we now describe.

\subsection{Estimation of $S_0(t)$ and $\delta(t)$}

We begin by defining additional notation pertinent to treatment-free data. Specifically, let $Y_i^0(t)=I(U_i\wedge T_i \geq t)$, an indicator for being at risk and untreated as of time $t$, and define the following counting process increment, $dN_i^0(t)=Y_i^0(t)dN_i(t)$. Suppose that patient $k$ is observed to initiate treatment at time $T_k$. We then let $I_{i:k}$ be an indicator for not-yet-treated subject $i$ being matched to patient $k$, with $I_{k:k}\equiv 0$. Since we apply one-to-one matching, we have $I_{\bullet:k}=\sum_{i=1}^nI_{i:k}$, consistent with the definition of $I_{\bullet:k}$ applied in (\ref{eq:w1k}). To specify matched-set-specific notation, we define $\widetilde{Y}_{i:k}^0(t)=Y_i^0(T_k+t)$ and $d\widetilde{N}_{i:k}^0(t)=dN_i^0(T_k+t)$.

As was the case for $S_1(t)$, we estimate $S_0(t)$ through a weighted version of the Nelson-Aalen estimator, with the
weight function for patient $i$ given by
\begin{eqnarray}
w_{i:k}^0(t) & = & \frac{N_k^T(\tau)I_{i:k}\widetilde{Y}_{i:k}^0(t)}{P(C_k>T_k|\bdZ_k,T_k)P(C_i \wedge T_i > T_k+t|C_i \wedge T_i>T_k, T_k, \bdZ_i)}. \label{eq:wik0}
\end{eqnarray}
The component $N_k^T(\tau)P(C_k>T_k|\bdZ_k,T_k)^{-1}$ is appropriately inherited from the weight assigned to
the treated patient.  A treated patient does not contribute to $\widehat{S}_1(t)$ unless $T_k<\tau$. Correspondingly,  patient $i$, matched to treated patient $k$, should not contribute to $\widehat{S}_0(t)$ unless patient $k$ was included.  With respect to the denominator, intuitively, inverse weighting patients $i$ and $k$ differently
at the time of matching ($t=T_k$) would serve to distort the balance in the covariates and previous-time-survived distributions achieved by matching. The gap time structure does not induce dependent censoring observed witnessed
on the post-treatment side. The pertinent gap times are $T_k$ and $(D_i^0-T_k)$. The latter is censored by $(C_i-T_k)$ but, unlike the analog on the $j=1$ side, $(D_i^0-T_k)$ can in fact be assumed to be independent of $T_k$.  Therefore, the
induced dependent censoring and associated identifiability issues do not naturally arise on the treatment-free side.
That said, we apply the afore-listed constraints to $S_0$ since it is desirable to contrast $S_1$ and $S_0$ functions which are truly comparable.

The third issue described in Subsection 2.3 is arguably a greater concern on the treatment-free side since, for matched patient $i$,
$(D_i^0-T_k)$ can be censored by either $(C_i-T_k)$ or $(T_i-T_k)$; both represent violations of independent censoring due to shared dependence on $\bdZ_i$, in the context of nonparametric survival curve estimation.
Note that our goal is to inversely weight the uncensored treatment-free experience with respect to the
distribution at $t=T_k$ as opposed to $t=0$; hence, the conditional probabilities.

The estimated version of (\ref{eq:wik0}) can be expressed as
\begin{eqnarray}
\widehat{w}_{i:k}^0(t) & = & N_k^T(\tau)I_{i:k}\widetilde{Y}_{i:k}^0(t)
\exp\left\{ \widehat{\Lambda}_{kC}(T_k)+ \int_{T_k}^{T_k+t}d\widehat{\Lambda}_{iC}(u) +\int_{T_k}^{T_k+t}d\widehat{\Lambda}_{iT}(u) \right\}, \nonumber \\
\label{eq:wik0hat}
\end{eqnarray}
where $\widehat{\Lambda}_{iT}(t) = \int_0^{t} \widehat{\lambda}_{iT}(u)du$.

Our proposed estimator of $S_0(t)$ is given by $\widehat{S}_0(t)=\exp\{-\widehat{\Lambda}_0(t)\}$, where
\begin{eqnarray}
 \widehat{\Lambda}_0(t)= \sum_{i=1}^n \sum_{k=1}^n \int_0^t \left\{ \sum_{\ell=1}^n \sum_{k=1}^n \widehat{w}_{\ell:k}^0(u) \right\}^{-1} \widehat{w}_{i:k}^0(u)d\widetilde{N}_{i:k}^0(u).
\end{eqnarray}
Note that in the treated group ($j=1$), every patient is unique but in the untreated group ($j=0$), the same patient can appear in multiple matched sets.

Having estimated $S_1(t)$ and $S_0(t)$, the ATT is then estimated by  $\widehat{\delta}(t)=\widehat{S}_1(t)-\widehat{S}_0(t)$.

\subsection{Variance Function}
The bootstrap (Efron 1979) is a frequently used method to estimate variances in settings such as ours where methods of analytically deriving the
variance are difficult. However,  Abadie and Imbens (2008) have shown that standard bootstrap is usually not valid for matching estimators. We propose a closed-form estimator that is convenient to compute. The aim is to arrive at a tractable form which captures the most important aspects of the
variability in the proposed estimators. We begin by considering the post-treatment side ($j=1$). Assuming that $\{Y_i(\cdot),N_i(\cdot),N_i^T(\cdot),\bdZ_i\}$ are independent and identically distributed for $i=1\ldots,n$, ignoring the randomness in the matching process and treating the weights as known, $n^{1/2}\{ \widehat{\Lambda}_1(t)-\Lambda_1(t) \}=n^{1/2}\sum_{i=1}^n \phi_i^1(t)$ asymptotically, where $\phi_i^1(t)=\int_0^t \pi_1(u)^{-1}w_i^1(u)d\widetilde{M}_i^1(u)$, with $\pi_1(u)=E[w_i^1(u)]$ and $d\widetilde{M}_i^1(u)=d\widetilde{N}_i^1(u)-\widetilde{Y}_i^1(u)d\Lambda_1(u)$  (e.g., see Andersen et al. 1993). Under mild regularity conditions,
the $\{\phi_1^1(t), \ldots, \phi_n^1(t)  \}$ are independent and identically distributed mean $0$ variates.  As a result, $n^{1/2}\{ \widehat{\Lambda}_1(t)-\Lambda_1(t) \}$ converges to asymptotically to a mean-zero Normal distribution with variance $E[\phi_i^1(t)^2]$, by the Multivariate Central Limit Theorem. By applying the Functional Delta Method (van der Vaart 2000), we obtain that $n^{1/2}\{ \widehat{S}_1(t)-S_1(t) \}$ is also asymptotically mean-zero Normal with variance estimator,
\begin{eqnarray}
\widehat{\sigma}_1^2(t) &= &n^{-1}\sum_{i=1}^n \{  \widehat{S}_1(t)\widehat{\phi}_i^1(t) \}^2,
\nonumber
\end{eqnarray}
where $\widehat{\phi}_i^1(t)=\int_0^t \widehat{\pi}_1(u)^{-1}\widehat{w}_i^1(u)d\widehat{M}_i^1(u)$, with $\widehat{\pi}_1(u)=n^{-1}\sum_{i=1}^n
\widehat{w}_i^1(u)$ and $d\widehat{M}_i^1(u)=d\widetilde{N}_i^1(u)-\widetilde{Y}_i^1(u)d\widehat{\Lambda}_1(u)$.  This is a robust version of the variance estimator and, as such, does not rely on Martingale theory (Fleming and Harrington 1991).

Then considering the treatment-free side ($j=0$), analogous arguments lead to $n^{1/2}\{ \widehat{S}_0(t)-S_0(t) \}=-n^{1/2}\sum_{i=1}^n S_0(t)\phi_i^0(t)$
asymptotically. Different from the treatment side where each subject $k$ can appear only once, a given subject $i$ in the treatment-free side can be matched to several treated patients. As such, the asymptotically independent
terms with respect to the treatment-free side are given by $\phi_i^0(t)=\sum_{k=1}^n \int_0^t \pi_0(u)^{-1}w_{i:k}^0(u)d\widetilde{M}_{i:k}^0(u)$,
and $n^{1/2}\{ \widehat{S}_0(t)-S_0(t) \}$ converges in distribution to a zero-mean Normal with a
variance that can be consistently estimated by
\begin{eqnarray}
\widehat{\sigma}_0^2(t)  & = & n^{-1}\sum_{i=1}^n \{  \widehat{S}_0(t)\widehat{\phi}_i^0(t) \}^2,
\nonumber
\end{eqnarray}
where we define
\begin{eqnarray}
\widehat{\phi}_i^0(t) & = &  \sum_{k=1}^n \int_0^t \widehat{\pi}_0(u)^{-1}\widehat{w}_{i:k}^0(u)d\widehat{M}_{i:k}^0(u) \nonumber \\
\widehat{\pi}_0(u) & = & n^{-1} \sum_{i=1}^n \sum_{k=1}^n \widehat{w}_{i:k}^0(u),
\nonumber
\end{eqnarray}
with $d\widehat{M}_{i:k}^0(u)=d\widetilde{N}_{i:k}^0(u)-\widetilde{Y}_{i:k}^0(u)d\widehat{\Lambda}_0(u)$.

Combining the above results, we can represent $n^{1/2}\{\widehat{\delta}(t)-\delta(t)\}$ asymptotically by  $n^{-1/2}\sum_{i=1}^n  \{S_0(t)\phi_i^0(t)-S_1(t)\phi_i^1(t)\}$,  where $S_0(t)\phi_i^0(t) - S_1(t)\phi_i^1(t)$ components are independent and identically distributed with mean $0$ and account for the possibility that patients may contribute follow-up on both the $j=0$ and $j=1$ sides. The quantity $n^{1/2}\{\widehat{\delta}(t)-\delta(t)\}$ converges asymptotically to a Normal variate with mean 0 and a
variance that can be consistently estimated by
\begin{eqnarray}
\widehat{\sigma}_\delta^2(t) & = & n^{-1}\sum_{i=1}^n \{  \widehat{S}_0(t)\widehat{\phi}_i^0(t) - \widehat{S}_1(t)\widehat{\phi}_i^1(t) \}^2.
\nonumber
\end{eqnarray}

The above-described variance estimators ignore the randomness in the matching process and the estimation of the weights. The idea of treating the weights as fixed for computational purposes is a commonly used simplification in the inverse weighting literature. In several analogous cases in the literature, it is argued that treating an estimated weight as known reduces precision (Hern$\mbox{\'{a}}$n, Brumback and Robins 2000 and 2001); heuristically, since the estimation is not credited for the extent to which it uses the data.  On the other hand, not accounting for matching process should result in under-estimation of asymptotic variances. However, our extensive simulations show that the randomness we ignored is usually small or almost negligible relative to the variability that is captured by $\widehat{\phi}_i^1(t)$ and $\widehat{\phi}_i^0(t)$ in even moderate sized samples.

\section{Simulations}

We conducted simulations to assess the performance of our proposed method in finite samples. The treatment times $T$ were generated from
an exponential distribution with hazard $\lambda_{0T}\exp\{\beta_{10} Z_1 + \beta_{11} Z_t \}$, while treatment-free death times were generated as
exponential with hazard $\lambda_{0D} \exp\{\beta_{20} Z_1 + \beta_{21} Z_d\}$. Note that the covariate $Z_t$ affects treatment assignment, but not death, while $Z_d$ has the opposite effect. The covariate $Z_1$ serves as a confounder that affects the rates of both treatment and treatment-free death. Times between treatment and death are generated from an exponential distribution with rate $\lambda_{1D}\exp\{\beta_{30} Z_1 + \beta_{31} Z_d + \beta_{32} \}$. Censoring times are generated from an exponential distribution with rate $\lambda_{0C} \exp\{ \beta_{40} Z_1\}$. Each of the covariates $Z_1$, $Z_t$ and $Z_d$ were generated from standard normal distributions. There were $n$=1,000 subjects in all simulations, with each data configuration replicated 1,000 times. In reality, we observe the minimum of the potential time to treatment, time to death and time to censoring. In simulations, however, we have every patient's potential time to treatment, time to death with and without treatment and time to censoring. As such, we can obtain the true effect of the treatment on survival by averaging the difference between the counterfactual survival functions across simulations.

In the first set of simulations, we examine the bias and empirical standard deviation of the proposed estimators using three different matching methods: (i) matching by prognostic score only (ii) matching by propensity score only (iii) matching by both prognostic and propensity scores; i.e., double matching. We do 1:1 nearest-neighbor within-caliper matching, with $\xi_T=1.1$ and/or $\xi_D=1.1$. Approximately $75\%$ treated patients find their matches in simulations. We first vary the magnitude of $\beta_{11}$ from $0$, $0.5$, $1$ to $1.5$ to change the degree of association between the predictor $Z_t$ and the treatment hazard from none, weak, moderate and strong, respectively.  We then vary the magnitude of $\beta_{21}$ from $0$, $0.5$, $1$ to $1.5$ to change the degree of association between the predictor $Z_d$ and the treatment-free death hazard. The remaining parameter specifications were equal across all simulations: $\lambda_{0T}=0.5, \lambda_{0D}=0.5, \lambda_{1D}=0.2, \lambda_{0C}=0.2, \beta_{10}= 0.15, \beta_{20}=0.25, \beta_{30} = 0.20, \beta_{31}=0.15, \beta_{32} = -0.7, \beta_{40} = 0.2$. We set $\tau=3$ and $\tau_1=5$. We present the average bias and the empirical standard deviation (ESD) of $\widehat{S}_1(t)$, $\widehat{S}_0(t)$ and $\widehat{\delta}(t)$ at $t=1.5$.

In the second set of simulations, we examine the properties of the proposed point estimators and their variance functions under four scenarios: (1) no treatment effect, where $\lambda_{0T}=0.7$ and $\lambda_{0D}=0.7=\lambda_{1D}=0.7$, $\beta_{20}=0.25, \beta_{21}=0.50, \beta_{30} = 0.20, \beta_{31}=0.50, \beta_{32} = 0$; (2) strong treatment effect, where $\lambda_{0T}=0.5, \lambda_{0D}=0.5, \lambda_{1D}=0.5, \beta_{20}=0.5, \beta_{21}=1, \beta_{30} = 0.20, \beta_{31}=0.15, \beta_{32} = -1$; (3) moderate treatment effect, where $\lambda_{0T}=0.5, \lambda_{0D}=0.5, \lambda_{1D}=0.7, \beta_{20}=0.25, \beta_{21}=0.5, \beta_{30} = 0.20, \beta_{31}=0.15, \beta_{32} = -0.7$; (4) negative treatment effect, where $\lambda_{0T}=0.5, \lambda_{0D}=0.5, \lambda_{1D}=0.7, \beta_{20}=0.25, \beta_{21}=0.5, \beta_{30} = 0.20, \beta_{31}=0.15, \beta_{32} = 0.4$. The other parameter specifications were the same across all scenarios: $\beta_{10} = 0.15, \beta_{11} = 0.5, \lambda_{0C}=0.2$ and $\beta_{40} = 0.2$. We conducted nearest-neighbor within-caliper prognostic score matching, with $\xi_D = 1.1$. We examined the proposed estimators (and their estimated standard errors) at $t=0.5$, $t=1$ and $t=1.5$.

As shown in Table $1$, the bias appears to be negligible for each of the proposed estimators. For $\widehat{S}_0(t)$, double matching method tends to give estimates with the smallest bias, followed by prognostic matching. For $\widehat{S}_0(t)$ and $\widehat{\delta}(t)$, prognostic score matching method tends to give much smaller empirical standard deviations, which are only $50\%$ to $80\%$ of those by either propensity score matching or double matching, except when $\beta_{11}=0$. Propensity score matching and double matching give rise to similar ESDs. Such results are consistent with the need to adjust for prognostic factors for the benefit of efficiency gain. As we increase the association between $Z_t$ and $T$ through $\beta_{11}$, both propensity score and double score matching methods tend to give increased empirical standard deviations. This illustrates that adjustment for a factor that is a stronger predictor of the treatment can actually lead to a increased variance if the factor is not a predictor of the outcome. For $\widehat{S}_1(t)$, differences in bias and empirical standard deviation between prognostic and propensity score matching methods are negligible in our simulations since almost all treated patients are able to find matches.

\begin{table}
\label{tab:1}
\caption{Bias and empirical standard deviation (ESD) of the estimates for survival functions for the untreated ($\hat{S}_0$) and treated ($\hat{S}_1$) patients, and their difference in survival ($\hat{\delta}$) at 1.5 years post treatment summarized across 1000 simulations where the association between $Z_t$ and $T$ ($\bbeta_{11}$) and that between $Z_d$ and $D^0$ ($\bbeta_{21}$) vary and three matching methods are considered}
\begin{center}
\begin{tabular}{lccccccccccccc}
\hline\noalign{\smallskip}
$\bbeta_{11}$ & $\bbeta_{21}$ & Matching & $\hat{S}_0$ 	&	Bias	&	ESD	 & $\hat{S}_1$ 	&	Bias	&	ESD &	 $\hat{\delta}$	&	Bias	&	ESD	 \\
\noalign{\smallskip}\hline\noalign{\smallskip}
0	&	1 & Prognostic	&	0.545	&	0.006	&	0.041	& 0.865	&	0.001	&	0.019	&	0.320	&	-0.006	&	 0.045 \\
0.5	&	1 &  	&	0.540	&	0.007	&	0.041	&	0.865	&	-0.000	&	0.019	&	0.325	&	-0.008	 &	 0.044 \\
1	&	1 & 	&	0.531	&	0.006	&	0.042	&	0.864	&	0.000	&	0.019 &	0.333	&	-0.006	&	 0.046 \\
1.5	&	1 & 	&	0.521	&	0.006	&	0.044	&	0.863	&	0.001	&	0.020 &	0.342	&	-0.005	&	 0.049	 \\
& & & & & \\
0 & 1	&	Propensity	&	0.545	&	0.006	&	0.042	&	0.865	&	0.001	&	0.019  &	0.320	&	-0.005	 &	0.046	\\
0.5 & 1	&		&	0.540	&	0.008	&	0.049	&	0.865	&	0.000	&	0.019	&	0.325	&	-0.008	 &	 0.052 \\
1 & 1	&		&	0.531	&	0.008	&	0.070	&	0.864	&	0.000	&	0.019	&	0.333	&	-0.008	 &	 0.072	 \\
1.5 & 1	&		&	0.521	&	0.013	&	0.089	&	0.863	&	0.001	&	0.020	&	0.342	&	-0.012	 &	 0.092	 \\
& & & & & \\
0 & 1	&	Double	&	0.545	&	0.003	&	0.042	&	0.865	&	0.001	&	0.020  &	0.320	&	-0.003	&	 0.046 \\
0.5 & 1	&		&	0.540	&	-0.002	&	0.049	&	0.865	&	-0.000	&	0.023	&	0.325	&	0.002	&	 0.054 \\
1 & 1	&		&	0.531	&	-0.002	&	0.065	&	0.864	&	0.000	&	0.024	&	0.333	&	0.002	&	 0.069 \\
1.5 & 1	&		&	0.521	&	0.004	&	0.087	&	0.863	&	0.000	&	0.026 &	0.342	&	-0.003	&	 0.091 \\
\noalign{\smallskip}\hline\noalign{\smallskip}
1	&	0 & Prognostic	&	0.472	&	0.000	&	0.046	&	0.859	&	-0.001	&	0.019  &	0.387	&	-0.001	 &	0.049 \\
1	&	0.5 & 	&	0.488	&	0.003	&	0.043	&	0.861	&	0.000	&	0.019  &	0.373	&	 -0.003	&	 0.046\\
1	&	1 & 	&	0.531	&	0.006	&	0.042	 &	0.864	&	0.000	&	0.019	&	0.333	&	-0.006	 &	 0.046 \\
1	&	1.5 & 	&	0.580	&	0.011	&	0.043	&	0.866	&	0.001	&	0.019	&	0.286	&	 -0.010	&	 0.047 \\
& & & & & \\
1 & 0	&	Propensity	&	0.472	&	0.005	&	0.070	&	0.859	&	-0.001	&	0.019 &	0.387	&	-0.006	&	 0.072 \\
1 & 0.5	&		&	0.488	&	0.007	&	0.072	&	0.861	&	0.000	&	0.019 &	0.373	&	-0.006	&	 0.075 \\
1 & 1	&		&	0.531	&	0.008	&	0.070	&	0.864	&	0.000	&	0.019 &	0.333	&	-0.008	&	 0.072 \\
1 & 1.5	&		&	0.580	&	0.012	&	0.069	&	0.866	&	0.001	&	0.019 &	0.286	&	-0.011	&	 0.072 \\
& & & & & \\
1 & 0	&	Double	&	0.472	&	0.004	&	0.067	&	0.859	&	-0.000	&	0.020 &	0.387	&	-0.004	&	 0.070 \\
1 & 0.5	&		&	0.488	&	0.003	&	0.064	&	0.861	&	0.001	&	0.021 &	0.373	&	-0.003	&	 0.067 \\
1 & 1	&		&	0.531	&	-0.002	&	0.065	&	0.864	&	0.000	&	0.024 &	0.333	&	0.002	&	 0.069	 \\
1 & 1.5	&		&	0.580	&	-0.006	&	0.067	&	0.866	&	-0.000	&	0.027 &	0.286	&	0.006	&	 0.071 \\
\noalign{\smallskip}\hline
\end{tabular}
\end{center}
\end{table}

The performances of the estimators of the survival functions and their variances in our simulations are summarized in Tables $2$ and $3$. We find that the biases are again negligible for all estimates of $S_0$, $S_1$ and $\delta$ at $0.5$, $1$ and $1.5$ years post treatment. The asymptotic standard errors are close to the empirical standard deviation of the estimates across all simulations, even when the estimates are relatively small (e.g., $\delta$, Table $2$). Both the biases and variances of the estimates tend to be larger at $t=1.5$ than those at $t=0.5$ because fewer subjects remain at-risk. Coverage probabilities are generally close to the nominal level of 0.95.

\begin{table}

\caption{Simulation summary of estimates of survival functions for the treated ($S_1$) and untreated ($S_0$) patients, and their difference ($\delta$) at time $t$ post treatment; Est: average of estimates of quantity of interest across $1000$ simulations; Bias: average bias across simulations; ESD: empirical standard deviation of estimates from $1000$ simulations; ASE: average of estimated standard errors across simulations; CP: coverage probability of nominal $95\%$ confidence interval; Null: no treatment effect; Strong: strong treatment effect.}
\label{tab:2}
\begin{center}
\begin{tabular}{lccccccc}
\hline\noalign{\smallskip}
Setting & $t$	&	Quantity	&	Est & Bias	&	ESD	&	ASE	&	CP	\\
\noalign{\smallskip}\hline\noalign{\smallskip}
Null & 0.5	& 	$S_0(t)$ 	& 	0.710	& 0.001 & 	0.027	& 	0.028	 & 	94.9	\\
& 1.0	& 	 	& 	0.519	& 0.001 & 	0.038	& 	0.038	 & 	95.0	\\
& 1.5	& 	 	& 	0.391	& 0.003 & 	0.044	 & 	0.045	& 	95.3	\\
& 	&		&		&		& &		&		\\
& 0.5	& 	$S_1(t)$ 	& 	0.711	& 0.003 & 	0.022	& 	0.023	 & 	94.6	\\
& 1.0	& 	 	& 	0.520	& 0.003 & 	0.026	& 	0.026	 & 	93.7	\\
& 1.5	& 	 	& 	0.389	& 0.003 & 	0.027	 & 	0.026	& 	93.7	\\
& 	&		&		&		& &		&		\\
& 0.5	& 	$\delta(t)$ 	& 	0.001	& -0.000 & 	0.035	& 	0.036	 & 	95.7	\\
& 1.0	& 	 	& 	0.001	& 0.002 & 	0.046	& 	0.046	 & 	94.3	\\
& 1.5	& 	 	& 	-0.002	& -0.001 & 	0.051	 & 	0.052	& 	94.6	\\
\noalign{\smallskip}\hline\noalign{\smallskip}
Strong & 0.5	& 	$S_0(t)$ 	& 	0.790	& 0.008 & 	0.023	& 	0.025	 & 	94.9	\\
& 1.0	& 	 	& 	0.652	& 0.008 & 	0.033	& 	0.034	 & 	93.8	\\
& 1.5	& 	 	& 	0.554	& 0.008 & 	0.040	 & 	0.041	& 	94.1	\\
& 	&		&		&		& &		&		\\
& 0.5	& 	$S_1(t)$ 	& 	0.916	& 0.001 & 	0.015	& 	0.015	 & 	93.8	\\
& 1.0	& 	 	& 	0.840	& 0.001 & 	0.020	& 	0.020	 & 	95.0	\\
& 1.5	& 	 	& 	0.769	& 0.001 & 	0.024	 & 	0.023	& 	94.3	\\
& 	&		&		&		& &		&		\\
& 0.5	& 	$\delta(t)$ 	& 	0.126	& -0.007 & 	0.027	& 	0.029	 & 	95.3	\\
& 1.0	& 	 	& 	0.187	& -0.007 & 	0.039	& 	0.040	 & 	94.8	\\
& 1.5	& 	 	& 	0.215	& -0.007 & 	0.046	 & 	0.048	& 	94.9	\\
\noalign{\smallskip}\hline
\end{tabular}
\end{center}
\end{table}

\begin{table}
\caption{Simulation summary of estimates of survival functions for treated ($S_1$) and untreated ($S_0$) patients, and their difference ($\delta$) at time $t$ post treatment; Est: average of estimates of quantity of interest across $1000$ simulations; Bias: average bias across simulations; ESD: empirical standard deviation of estimates from $1000$ simulations; ASE: average of estimated standard errors across simulations; CP: coverage probability of nominal $95\%$ confidence interval; Medium: medium treatment effect; Negative: negative treatment effect.}
\label{tab:3}
\begin{center}
\begin{tabular}{lccccccc}
\noalign{\smallskip}\hline
Setting & $t$	& Quantity	&	Est & Bias	&	ESD	&	ASE	&	CP	\\
\noalign{\smallskip}\hline\noalign{\smallskip}
Medium & 0.5	& 	$S_0(t)$ 	& 	0.778	& 0.002 & 	0.025	& 	0.025	 & 	94.5	\\
& 1.0	& 	 	& 	0.617	& 0.004 & 	0.035	& 	0.035	 & 	93.6	\\
& 1.5	& 	 	& 	0.495	& 0.002 & 	0.041	 & 	0.042	& 	94.7	\\
& 	&		&		&		& &		&		\\
& 0.5	& 	$S_1(t)$ 	& 	0.841	& 0.001 & 	0.019	& 	0.019	 & 	95.4	\\
& 1.0	& 	 	& 	0.709	& 0.001 & 	0.024	& 	0.024	 & 	95.1	\\
& 1.5	& 	 	& 	0.599	& 0.001 & 	0.027	 & 	0.027	& 	95.6	\\
& 	&		&		&		& &		&		\\
& 0.5	& 	$\delta(t)$ 	& 	0.063	& -0.001 & 	0.031	& 	0.031	 & 	94.5	\\
& 1.0	& 	 	& 	0.092	& -0.003 & 	0.042	& 	0.042	 & 	95.2	\\
& 1.5	& 	 	& 	0.104	& -0.001 & 	0.049	 & 	0.049	& 	95.1	\\
\noalign{\smallskip}\hline\noalign{\smallskip}
Negative & 0.5	& 	$S_0(t)$ 	& 	0.777	& 0.001 & 	0.026	& 	0.025	 & 	94.4	\\
& 1.0	& 	 	& 	0.614	& 0.001 & 	0.034	& 	0.035	 & 	95.2	\\
& 1.5	& 	 	& 	0.495	& 0.002 & 	0.042	 & 	0.041	& 	95.2	\\
& 	&		&		&		& &		&		\\
& 0.5	& 	$S_1(t)$ 	& 	0.599	& 0.002 & 	0.026	& 	0.025	 & 	93.4	\\
& 1.0	& 	 	& 	0.364	& 0.001 & 	0.026	& 	0.026	 & 	95.2	\\
& 1.5	& 	 	& 	0.225	& 0.000 & 	0.024	 & 	0.023	& 	93.7	\\
& 	&		&		&		& &		&		\\
& 0.5	& 	$\delta(t)$ 	& 	-0.178	& 0.003 & 	0.036	& 	0.036	 & 	93.4	\\
& 1.0	& 	 	& 	-0.250	& -0.001 & 	0.042	& 	0.043	 & 	95.7	\\
& 1.5	& 	 	& 	-0.270	& -0.002 & 	0.048	 & 	0.047	& 	95.1	\\
\noalign{\smallskip}\hline
\end{tabular}
\end{center}
\end{table}

\section{Application}

We applied the proposed methods in order to estimate the effect of deceased-donor kidney transplantation ($j=1$) on survival compared to dialysis ($j=0$) among end-stage renal disease patients. Data were obtained from the Canadian Organ Replacement Register, a nation-wide and population-based organ failure registry.  The study population included $n$=27,424 patients aged $\geq$18 years who initiated dialysis in Canada between 1989 and 1998. Patients began follow-up at the date of dialysis initiation and were followed until the earliest of death, loss to follow-up, or the end of the observation period (December 31, 1998).  Adjustment covariates for each of the death-on-dialysis, transplant, and censoring hazard models, included age, sex, race, province, diagnosis, calendar year of therapy initiation, initial dialytic
modality, and number of comorbid conditions.

We set $\tau$ and $\tau_1$ to 3 years and 5 years, respectively.  A total of
3,135 patients received a deceased-donor kidney transplant within the first 3 years of follow-up. We performed prognostic score matching, with $\xi_T=\xi_D=1.05$.  Under this matching scheme, all 3,135 transplants were matched.  There were 619 observed deaths on dialysis and 460 post-transplant deaths.

Figure $2$ shows $\widehat{S}_1(t)$, the average survival curve from the time of transplant among patients transplanted, and $\widehat{S}_0(t)$ intended to represent the survival curve (again, following transplant) to which the transplanted patients would have been
subjected had kidney transplantation been unavailable. Examining $\widehat{S}_1(t)$, 1-, 3- and 5-year survival is estimated at 0.95, 0.90 and 0.85, respectively. In contrast, in the absence of kidney transplantation, 1-, 3- and 5-year survival is estimated to be 0.92, 0.74 and 0.60. The lack of proportionality between the $\widehat{\Lambda}_1(t)$ and $\widehat{\Lambda}_0(t)$ is examined more closely in the right panel in Figure 2 when we magnify $\widehat{S}_1(t)$ and $\widehat{S}_0(t)$ such that only the $(0,1]$ year interval is displayed. The crossing of the survival functions is apparent from this plot.

\begin{figure}
\begin{center}
\centerline{\includegraphics[width=17.4cm, height=15cm]{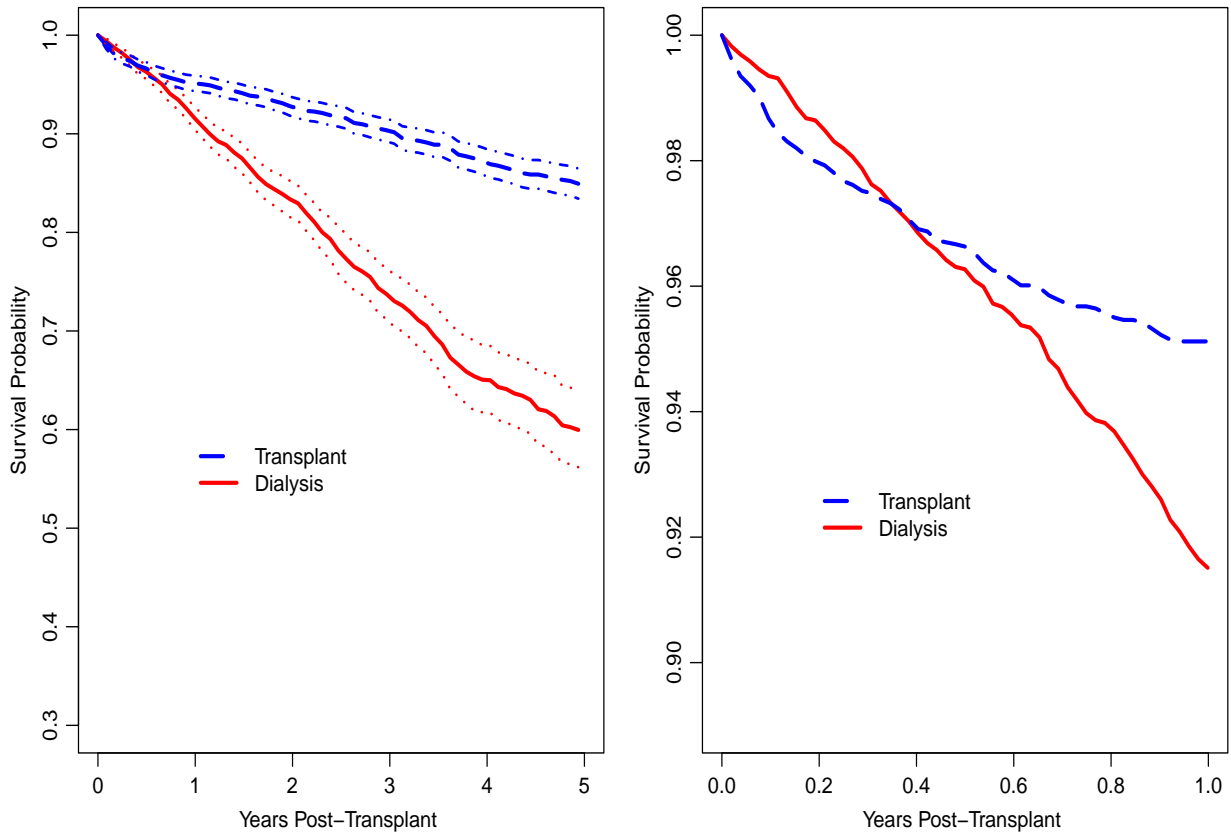}}
\end{center}
\caption{Post-transplant average survival curves for dialysis and transplanted patients. Left: Estimates and $95\%$ confidence intervals for 5-year post-transplant survival curves; Right: Estimates for 1-year post-transplant survival curves
\label{fig:2}}
\end{figure}

The proposed treatment effect estimator, $\widehat{\delta}(t)=\widehat{S}_1(t)-\widehat{S}_0(t)$, is presented in Figure 3.
For the first 4 months, it is estimated that survival is actually higher on dialysis, $\widehat{\delta}(t)<0$, due to the mortality risk
associated with surgery not faced by patients continuing dialysis.  From $t$=5 months on, we estimate $\widehat{\delta}(t)>0$, the contrast being
significant from $t$=9 months on.

\begin{figure}
\begin{center}
\centerline{\includegraphics[width=17.4cm, height=15cm]{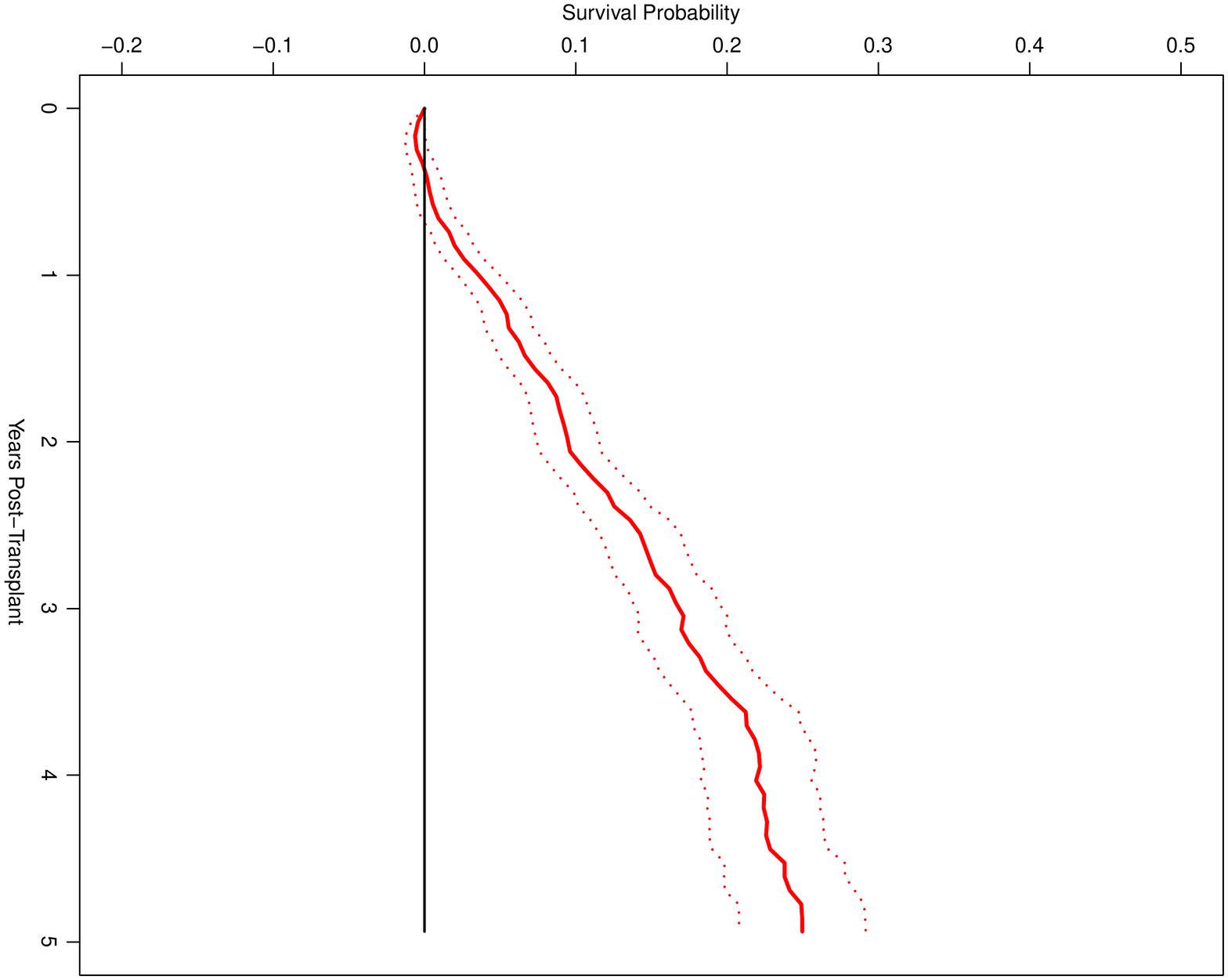}}
\end{center}
\caption{Estimate and $95\%$ confidence interval for difference between post-transplant average survival curves for dialysis and transplanted patients
\label{fig:3}}
\end{figure}

\section{Discussion}

In this report, we developed matching methods to estimate pertinent survival functions used in turn to estimate the average causal effect on the treated of a time-dependent treatment. In particular, the proposed treatment effect compares post-treatment survival with the survival experience to which the treated patients would have been subjected in the absence of treatment. The ATT is particularly interesting for policy makers in evaluating the realized impact of treatment implementation given the observed treatment assignment patterns.  The proposed methods do not require that
pre- and post-treatment death hazards are proportional.  Analytical forms of the variances are proposed and shown through simulations to work well in practical samples sizes.  Our methods are easy to implement in SAS or R and the code can be downloaded at https://github.com/yunliyunli/Matching.

Several existing methods are related to those proposed. Generally, such methods do not do one or more of the following: target the survival function; estimate the ATT; use time-from-treatment as the time scale. Marginal structural models (Robins et al. 2000; Hern$\mbox{\'{a}}$n et al. 2000 and 2001; Petersen et al. 2007) usually target the causal hazard ratio, which generally cannot be used to obtain survival functions due to the nature of the averaging.  The accelerated failure time model assumed in g-estimation (Robins et al. 1992; Lok et al. 2004; Hern$\mbox{\'{a}}$n et al. 2005) measures ratios of mean survival times, as opposed differences between survival functions. Unlike the proposed methods, g-estimation typically involves parameterizing the treatment effect.  The treatment effect (a time scale acceleration factor) is typically represented as a one-number summary, implicitly assuming that the treatment effect is equal for all patients. The treatment effect can be parameterized more generally, but at increased computational expense since a estimation requires a grid search.  Parametric g-computation (Robins 1986, 1987 and 1988; Taubman et al. 2009) could be used to estimate the contrast in survival functions estimated by the proposed methods. However, unlike the proposed methods, a post-treatment survival model would be required. In contrasting our methods with the g-formula, the latter would likely be more sensitive to model misspecification (Taubman et al. 2009); i.e., since the models are used to actually simulate data, as opposed to classify or weight subjects. Additionally, matching can handle high dimensional covariates (Rosenbaum and Rubin 1983). Conversely, g-computation could offer
 efficiency gains relative to the proposed methods, if the assumed models were all correct. With respect to computational
 convenience, the g-formula would require the bootstrap, which is cumbersome for large data sets; e.g., the CORR database we analyzed in Section 4.

Since we consider time-constant covariates, the data structure of our interest can be considered a special case of that dealt with in the
methods listed in the preceding paragraph.  It would appear that our methods could readily be extended to
handle time-dependent covariates. It seems that our methods would carry through as detailed in Section 2, if the assumed models for each of censoring, treatment and pre-treatment death were modified to incorporate time-dependent covariates.  Our interest was not in this area since, like many large registry-based data sets, time-dependent covariates were not present in the motivating example.

Each of the three different matching methods we considered (propensity score matching, prognostic score matching and double matching) yields unbiased treatment effect estimators in our simulations. However, prognostic score matching appears to provide the most efficient estimators, which reflects the importance to precision of adjusting for imbalances in covariates associated with mortality. On the other hand, we observed that adjusting for covariates associated only with treatment can actually increase the variance. Such findings are consistent with those of Chen and Tsiatis (2001).

Propensity score matching was first proposed by Rosenbaum and Rubin (1983), with its existing applications having mostly being limited to settings wherein treatment is assigned at baseline. When treatment assignment is dynamic, as in a longitudinal observational study, Li, Propert and Rosenbaum (2001) proposed a balanced risk set matching design, while Lu (2005) proposed a time-dependent propensity score. However, these methods do not deal with time-to-event outcomes or matching with replacement. Prognostic score matching has been discussed by Hansen (2008) when the treatment is assigned at baseline and is similar to the predicted mean matching in missing data literature (Little and Vartivarian 2005; Hsu and Taylor 2011). Time-dependent prognostic score matching has been used by Prentice and Breslow (1978) for a case-control study. Other work on time-dependent matching includes that by Abbring and Van Der Berg (2004), Fredriksson and Johansson (2008), Schaubel, Wolfe and Port (2006), and Schaubel et al (2009).

We used within-caliper nearest-neighbor matching with replacement, which allows every treated patient to be matched to the yet-untreated patient with the closest score, even if that match has been used previously for other treated patients at previous different times. Matching with replacement reportedly selects closer matches, produces more efficient and less biased estimates and is less sensitive to the order of the matches, compared to matching without replacement (Dehejia and Wahba 2002). In this paper, we conduct one-to-one matching. However, there may be efficiency advantages to selecting multiple matches per treated patient.

We develop analytic techniques for estimating the variance of the proposed estimators.  The variance estimators account for the correlations arising from the matching-with-replacement process, although the randomness in the matching and inverse weighting processes is ignored. In our simulations, the variance estimates were very close to the true values and the coverage probabilities were close to the nominal level.
Such approximations are attractive since popular resampling methods (e.g., the bootstrap) often fail to yield valid variance estimators in the presence of nearest-neighbor matching (Abadie and Imbens 2008). Although the randomness introduced by matching and weighting is often ignored (e.g., Stuart 2010), it would be worthwhile to further investigate the impact of these considerations. Ideally, standard errors which accounted for all sources of variation would be preferred, provided that the result is computationally convenient.

The matching methods proposed here relies on the assumptions such as no unmeasured confounding, overlapping support between treated and untreated groups, no measurement errors and no model misspecification (Rosenbaum and Rubin 1983). We also assume stable unit treatment value assumption and strong ignorability of the treatment assignment given the propensity (or prognostic or double) scores (Rubin 1974 and 1976; Rosenbaum and Rubin 1983). Our quantity of interest is ATT which is redefined for the common support and can be more meaningful in practice when there is a self-selection among the treated patients. The finite sample positivity violations and the selection of common support impact our methods in terms of how large the caliper needs to be made to ensure subjects are not thrown out from failure to find a match. While we choose a caliper for simplicity, other choices of caliper are worthy of exploration.

\begin{acknowledgements}
The authors would like to thank Jeffrey Smith and Brenda Gillespie for their valuable input
on the work.

\end{acknowledgements}


{}

\end{document}